# On Crystal Lattice Parameters of Graphite-Like Phases of the B–C System

O. O. Kurakevych[a, b], T. Chauveau[a], and V. L. Solozhenko[a]

[a] IMPC, Université P&M Curie, Paris, France
[b] LPMTM-CNRS, Université Paris Nord, Villetaneuse, France

**Abstract**—The structure of graphite-like $BC_x$ phases ($x = 1, 1.5, 3, 4, 32$) has been studied using conventional X-ray diffraction. The results have been obtained, which unambiguously point to turbostratic (one-dimensionally disordered) structure of all phases under study. The crystal lattice parameters, sizes of coherent scattering domains, and microstrain values have been defined, which have allowed us to find a correlation between the structure and stoichiometry of the phases synthesized at the same temperature.

## I. INTRODUCTION

Carbon and boron are the elements that have the hardest allotropic modifications [1, 2] and possess the unique set of physicochemical properties [3–5]. Until recently, of compounds of these elements only boron subcarbides $B_4C$ [6], $B_{50}C_2$ [7], and graphite-like $BC_x$ ($x = 1.5–19$) phases were known [8]. The recent synthesis of diamond-like $BC_5$ [9] and obtaining superhard conducting composites based on phases of the B–C system [10, 11] have shown that graphite-like $BC_x$ phases offer promise as the initial materials to produce novel superhard materials and as a consequence have attracted the attention of researchers to the studies of these phases under pressure [12–15]. However, many properties of graphite-like phases of the B–C system remain unstudied even under ambient conditions. At present there is no consensus of notions of the structure of boron-doped graphite-like layers as well. The suggested models include both the random and ordered substitutions of boron atoms for carbon atoms in flat graphene layers [8–16]. Because of the fact that atomic factors of carbon and boron scattering are close, to define the relative arrangement of boron and carbon atoms in a layer from the intensities of diffraction lines seems impossible.

The attempts to study the effect of the composition on the lattice parameters and structure defectiveness of graphite-like phases of the B–C system were made earlier in [8, 17]. However, the calculation of an interlayer spacing was of an evaluation type (from the position of the single 001 line), the $a$ parameter of the two-dimensional crystal lattice was never defined, and the real structure of the phases was described in terms of full-width at half-maximum (FWHM), which is a complex function both of the coherent scattering domain size and of lattice microstrain [18], and hence, cannot be an unambiguous characteristic of the defect concentration in a structure. Nevertheless, a nonmonotonic dependence of interlayer spacings on the boron concentration in a graphite-like lattice has been established and shown that the interlayer spacing depends not only on the composition but on the synthesis temperature as well [8, 17].

In this paper we report the results of the precision determination of the lattice parameters of graphite-like phases of the B–C system and of turbostratic graphite (tC) produced at 1500 K. The real structure of these phases was characterized in terms of the linear size of coherent scattering domains and the lattice relative microstrain.

## 2. EXPERIMENTAL AND CALCULATION METHODS

Graphite-like $BC_x$ phases ($x = 1, 1.5, 3, 4, 32$, and $\infty$) were synthesized following the procedure described in [8] by depositing products of the interaction between acetylene, boron trichloride, and hydrogen at 1500 K onto a substrate

The structure was studied by X-ray diffractometry on a Seifert MZIII automated diffractometer ($CuK\alpha_{1,2}$ radiation, $\langle\lambda\rangle = 1.54187$ Å) in Bragg-Brentano geometry. The goniometer was adjusted using the $LaB_6$ stan-dard sample (space group Pm-3m, $a = 4.15695$ Å [19]). Diffraction patterns were taken with the step-size of 0.04° (2θ), the time of accumulation being 30–120 s. The use of thin-plate samples (~0.1 mm thick) has allowed us to avoid errors caused by absorption of X-rays by a sample.

Positions and broadenings (FWHM, the full-width at half-maximum) of the $00l$ symmetrical lines in diffraction patterns were defined by fitting the line profile to Pearson's function using the convergence method with the help of the DatLab program. The positions of the $hk0$ asymmetrical lines and the sizes of coherent scattering domains were determined by fitting the line profile to the Warren function [20]

$$I = A \cdot \frac{1 + \cos^2 2\theta}{2(\sin\theta)^{3/2}} \cdot \left(\frac{L_a}{\sqrt{\pi}\lambda}\right)^{1/2} F(a), \quad (1)$$

where $I$ is the intensity of the scattered radiation, $a = (2\sqrt{\pi}L/\lambda)\cdot(\sin\theta - \sin\theta_0)$, $L_a$ is the size of a coherent scattering domain along the $a$ and $b$ crystallographic axes, $\theta_0$ is the peak position, $\lambda$ is the wavelength used, $A$ is the proportionality coefficient and $F(a) = \int_0^\infty \exp[-(x^2 - a^2)]dx$. The experimental profile of the line was fitted to Eq. (1) using a simplex technique in the search of solutions (direct search) realized with the help of the MatLab program.

The corrections for zero shift $z_0$ (in 2θ units) and sample displacement from diffraction plane $d_0$ (in 2θ units) were taken into account by the $2\theta = 2\theta_{exp} - z_0 - d_0 \cdot \cos\theta$ formula. The $z_0$ value was found from the line displacements of standard sample $LaB_6$, while the $d_0$ parameter was defined together with the lattice parameters of the phases under study using the U-Fit program.

Sizes of the coherent scattering domains, $L_c$, and microstrain, $\varepsilon_c$, in the direction of the $c$ axis were calculated by the Wagner–Aqua formula [18], which relates the broadening of the line and its angle position (2θ) in the diffraction pattern:

$$\frac{(\delta 2\theta)^2}{\tan^2\theta_0} = \frac{K\lambda}{L_c}\left(\frac{\delta 2\theta}{\tan\theta_0 \sin\theta_0}\right) + 16\varepsilon_c^2, \quad (2)$$

where $K = 0.91$. This model supposes that a decrease in the sizes of the coherent scattering domains results in the broadening of the Lorentz component of the line profile, while an increase in the fraction of microstrains leads to the broadening of the Gaussian component of the line profile [18]. The parameters of Eq. (2) were calculated using the least-squares method (MatLab). As the $\delta 2\theta$ broadenings, the experimental FWHM values ($\delta_{exp} 2\theta$) with allowance made for corrections by the formula $\delta 2\theta = \delta_{exp} 2\theta - \delta_{instr} 2\theta - \delta_{K\alpha} 2\theta$, where $\delta_{instr} 2\theta$ is the instrumental broadening defined using the $LaB_6$ standard sample, and $\delta_{K\alpha}$ is the broadening, which for wide reflections of turbostratic phases under study is equal to the value of the splitting (in the 2θ units) of the corresponding interlayer reflection into a doublet due to the existence of two wavelengths ($CuK\alpha_1$ and $CuK\alpha_2$).

To compare with the results of previous studies, the $L_c$ value was evaluated by the Sherer equation [18],

$$\delta 2\theta = \frac{K\lambda}{L_c \cos\theta}. \quad (3)$$

## 3. RESULTS AND DISCUSSION

The crystal structure of graphite-like phases of the B–C system is well described by a model of turbostratic (one-dimensionally disordered) random layer lattice suggested by Warren [20]. In the framework of this model a crystal lattice is characterized by two parameters: the $a$-parameter of the two-dimensional crystal lattice of the layer and the $c$-parameter that corresponds to interlayer spacing. Because of the absence of a correlation in mutual arrangement of atoms between various layers, only the $hk0$ reflections of individual layers and $00l$ interlayer reflections are observed in diffraction patterns of turbostratic structures. All $hkl$ reflections, for which $|h| + |k| \neq 0$, $l \neq 0$, exhibit zero intensity [20].

The $c$ lattice parameter was calculated from symmetric lines 001 and 002 (Table 1), while the $a$ parameter—from asymmetric lines 10 and 11 (Table 2). Diffraction patterns of turbostratic phases $BC_x$ of compositions $x = 3, 4, 32$ and turbostratic graphite ($x = \infty$) indicate that the phases are heavily textured, which follows from a clear dependence of relative intensities of the $00l$ and $hk0$ lines on the sample orientation in an X-ray beam (Fig. 1b). In this case, the $a$ and $c$ parameters were calculated based on two patterns taken when the sam-ple was differently oriented with respect to the diffraction plane. A random texture was observed in the case of high boron content samples ($x = 1$ and $1.5$) (Fig. 1a), which made possible the calculation of both lattice parameters from the same diffraction pattern. The asymmetric profiles of the $hk0$ reflections showed a good fit to the Warren function [20] (Fig. 2), while symmetric profiles of the $00l$ lines fitted to the Pearson function.

**Table 1.** Parameters $c$ of the crystal lattice, sizes of coherent scattering domains $L_c$ and structure relative microstrains $\varepsilon_c$ of turbostratic phases of the B–C system

| Phase | 2θ (001; 002) | FWHM (001; 002) | c | $d_0$ | Sherrer $L_c$, Å | Wagner–Aqua $L_c$, Å | Wagner–Aqua $\varepsilon_c$, % |
|---|---|---|---|---|---|---|---|
| BC | 25.967; 53.674 | 1.47; 1.57 | 3.4031 | −0.16 | 66[1]; 76[2] | ~63 | ~0.40 |
| $BC_{1.5}$ | 25.674; 53.016 | 1.46; 2.72 | 3.4428 | −0.14 | 67[1]; 38[2] | 166 | 1.77 |
| $BC_3$ | 25.650; 53.047 | 1.31; 2.63 | 3.4373 | −0.21 | 76[1]; 40[2] | 359 | 1.84 |
| $BC_4$ | 26.086; 53.532 | 0.78; 1.49 | 3.4286 | 0.17 | 148[1]; 66[2] | 450 | 0.86 |
| $BC_{32}$ | 26.109; 53.878 | 1.40; 2.75 | 3.3956 | −0.07 | 70[1]; 41[2] | 139 | 1.55 |
| C | 25.471; 52.908 | 4.08; 5.06 | 3.4347 | −0.41 | 21[1]; 16[2] | 24 | 2.75 |

Notes: [1] calculated from broadening of line 001; [2] calculated from broadening of line 002.

**Table 2.** Parameters $a$ of the crystal lattice and sizes of coherent scattering domains $L_a$ of turbostratic phases of the B–C system

| Phase | 2θ (10; 11) | $d_0$ | a | Warren $L_a$, Å |
|---|---|---|---|---|
| BC | 42.341; – | −0.16[1] | 2.4531[2] | 15[2] |
| $BC_{1.5}$ | 42.422; – | | 2.4535[2,4] | 14[2,4] |
| $BC_3$ | 42.235; 77.143 | 0.13 | 2.4738 | 15[5], 9[3] |
| $BC_4$ | 42.416; 77.084 | 0.51 | 2.4834 | 13[2] |
| $BC_{32}$ | 42.267; 77.375 | −0.01 | 2.4645 | 6.2[2]; 4.9[3] |
| C | 42.665; 78.198 | −0.01 | 2.4427 | 5.8[2]; 4.5[3] |

Notes: [1] The parameter was calculated based on the positions of lines 001 and 002; [2] the parameter was calculated from the profile of line 10; [3] the parameter was calculated from the profile of line 11; [4] the parameter was calculated from the data obtained by energy dispersive diffractometry with synchrotron radiation; [5] the parameter was found by the trial-and-error method.

The calculated values of the $a$ parameter of the graphene layer two-dimensional crystal lattice and interlayer spacing $c$ are given in Tables 1 and 2. Average linear sizes of the coherent scattering domains along the $a$ axis makes $L_a = 5–15$ Å, which corresponds to 2–20 hexagons of B and C atoms. It should be noted that $L_a$ increases with the boron content and attains a constant value (~ 15 Å) in the region of compositions $x = 3–4$. The values of the linear sizes of the coherent scattering domains along the $c$ axis ($L_c$) calculated by Sherer equation (3) turned out to be not as unambiguous: the values calculated from the broadening of the 001 lines are approximately twice as high as the corresponding values calculated from the broadening of the 002 lines. The similar effect was observed for turbostratic graphite even in Franklin's earlier studies [21]. Obviously, it stems from the fact that the line broadening is caused not only by the size of the coherent scattering domains (as is suggested in the framework of the Sherer equation, which is applicable only in the case of the ideal structure or extensive defects that are boundaries of crystallites [18]), but also by the structure microstrains (point defects or systematic displacement of atom groups from the ideal positions in the lattice).

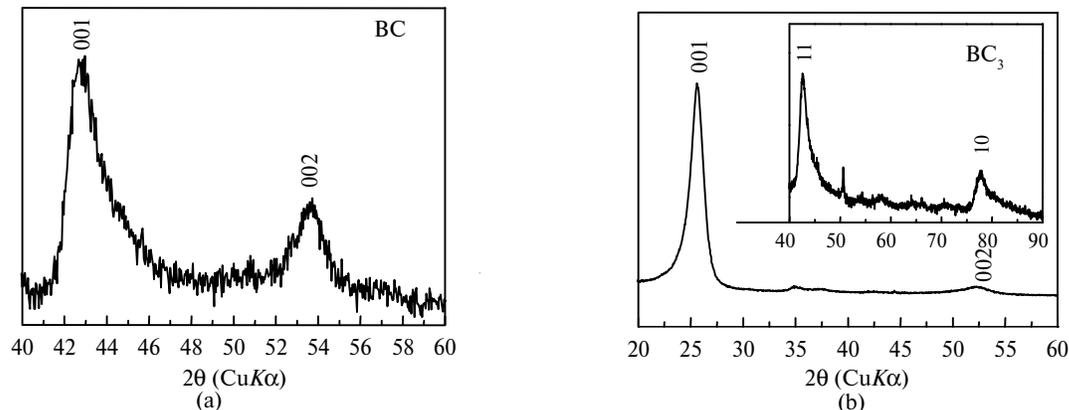

**Fig. 1.** Diffraction patterns of the tBC sample (a) and of the $tBC_3$ textured sample (diffraction patterns in the figure and in the inset correspond to mutually perpendicular orientations of a sample with respect to the diffraction plane) (b).

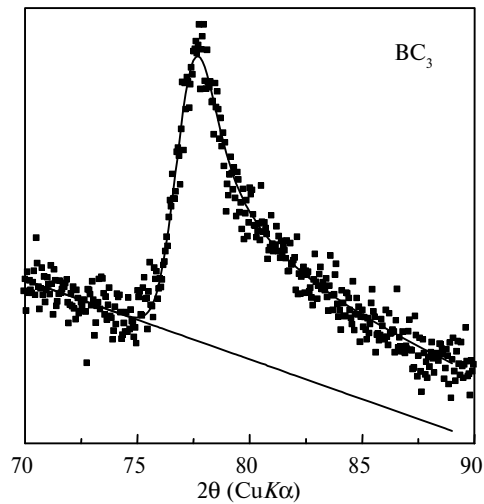

**Fig. 2.** Line 11 in the diffraction pattern of the tBC$_3$ sample (squares indicate experimental points; solid line shows the result of fitting the line profile to Warren's function).

The $L_c$ values calculated by Eq. (2) are at their maximum when the boron content corresponds to $x = 3-4$. The structure relative strain value along the $c$ axis is $= 1-3\%$ and has its minimum in the same concentration region. This behavior of the $L_c$ and $\varepsilon_c$ dependences on the composition is indicative of the existence of phases having a relatively ordered interlayer structure in the region of compositions $x = 4-5$, while a strong (almost triple) increase of the $L_a$ value with increasing boron concentration indicates that boron exerts the ordering effect on graphene layers.

Parameters of the crystal lattice of turbostratic phases BC$_x$ of various compositions are shown in Fig. 3. The concentration dependence of the $a$ parameter is of the extremum nature with a maximum at ~15 at % boron (the composition is close to the stoichiometry of BC$_5$). The concentration dependence of the $c$ parameter is of the opposite nature: as the boron concentration increases, the initial decrease in the interlayer spacing is followed by an increase with a maximum at the composition BC$_{1.5}$. This dependence of the $c$ parameter on the composition points to a strong influence of boron atoms embedded in graphite layers on the structure of the forming phases. Thus, unlike the ordered graphite, in which the embedding of boron atoms into the lattice results in a decrease of the *c* parameter [22], for turbostratic phases an increase of the boron concentration results in a directly opposite effect, which may be attributable either to the weakening of van der Waals bonds between doped graphene layers (the covalent radius of boron is too large to suppose the intercalation of boron atoms), or to a partial puckering of layers.

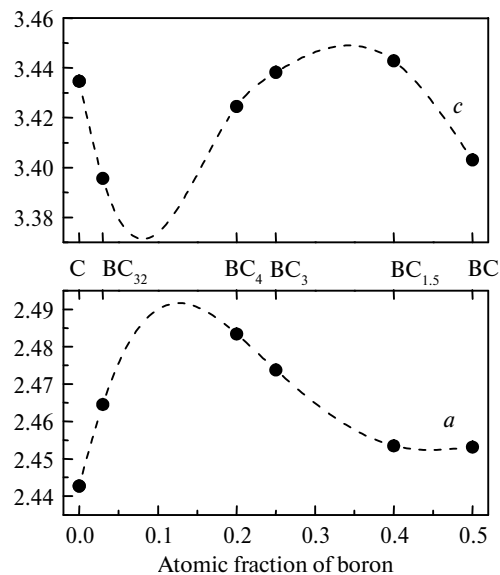

**Fig. 3.** Parameters of crystal lattice of turbostratic phases BC$_x$ vs. boron concentration.

The concentration dependence of structural parameters established in the present study is evidently universal for series of graphite-like phases $BC_x$ produced at the same temperature. Similar concentration dependences were observed, e.g., for bulk modulus of graphite-like phases of the B–C system [15].

## 4. CONCLUSIONS

X-ray diffractometry has been used to define parameters of crystal lattice, linear sizes of the coherent scattering domains, and the structure microstrain of turbostratic graphite-like phases of the B–C system produced at the same temperature and having different compositions. It has been shown that the corresponding concentration dependences are of clearly defined nonmonotonic nature.

The authors thank Dr. Derré for the kindly supplied samples of graphite-like phases of the B–C system and the Agence Nationale de la Recherche for the financial support (grant ANR-05-BLAN-0141).